\documentclass[letterpaper,10pt]{article} 
\usepackage{opticameet3} 

\newcommand\authormark[1]{\textsuperscript{#1}}

\usepackage{amsmath,amssymb}
\usepackage[colorlinks=true,bookmarks=false,citecolor=blue,urlcolor=blue]{hyperref}
\usepackage{blindtext}
\usepackage{array}
\usepackage{multirow}
\usepackage[normalem]{ulem}
\usepackage[countmax]{subfloat}
\usepackage{fancyhdr}
\usepackage{subcaption}

\usepackage{comment}
\usepackage{subfloat}
\usepackage{graphicx}
\graphicspath{ {Images/} }
\usepackage{enumitem}
\setlist[itemize]{align=parleft,left=0pt..1em}

\begin{document}

\pagestyle{fancy}
\fancyhf{}
\fancyhead{} 
\renewcommand{\headrulewidth}{0pt} 
\fancyfoot[C]{\footnotesize This paper is a preprint of a paper accepted to OFC 2025 and is subject to OFC 2025 copyright.}
\title{Coexistence of Digital Coherent, mmWave and sub-THz Analog RoF Services using OSaaS over converged access-metro live production network\vspace{-6mm}}



\author{
    Devika Dass \authormark{(1)}, Amol Delmade \authormark{(2)},
    Agastya Raj \authormark{(1)}, Eoin Kenny \authormark{(3)}, Dan Kilper \authormark{(1)},  Liam Barry \authormark{(2)} and Marco Ruffini \authormark{(1)}}

\maketitle                  


\address{
  \textsuperscript{1} Schools of Computer Science / Engineering, CONNECT, Trinity College Dublin, Ireland\\ 
  \textsuperscript{2} School of Electronic Engineering, Dublin City University, Ireland,
  \textsuperscript{3} HEAnet CLG, Dublin, Ireland
}
        
        \email{\href{mailto:dassd@tcd.ie}{\textcolor{blue}{dassd@tcd.ie}},
        \href{mailto:marco.ruffini@tcd.ie}{\textcolor{blue}{marco.ruffini@tcd.ie}}}

\renewcommand\footnotemark{}
\renewcommand\footnoterule{}

    \vspace{-6mm}
    \begin{abstract}
We demonstrate the end-to-end transmission of digital coherent and analog radio-over-fiber signals, at mmWave and sub-THz frequencies, over the HEAnet live production metro network using Optical Spectrum-as-a-Service (OSaaS), transparently connected to a passive optical network.
    \end{abstract}
    \vspace{-1mm}


\section{Introduction}
\vspace{-2mm}
5G and its evolution toward 6G networks expand beyond the target performance of higher capacity and lower latency. The aim is towards customization, with requirements like energy efficiency, reliability, and cost per connection, hence the optical networks must evolve to provide flexibility across multiple dimensions. Optical networks must connect service generation points, such as data centers or edge computing nodes, to ubiquitous locations across cities and suburban areas, to support wireless services close to end users \cite{5G_sharing, zhu2024optical}. The shift to higher wireless frequencies is being exploited for larger capacity, as more radio units (RU) are deployed at the access sites due to the higher propagation loss. This ignites the need for technologies assisting low-cost RUs, centralized signal processing and network resource sharing. Additionally, connectivity must be flexible to optimize component multiplexing across a wide array of services, in a spectrally efficient manner to maximize fiber utilization, especially as we move towards Optical Spectrum as a Service (OSaaS) solutions.

This work demonstrates, for the first time, the integration of essential technologies in a field trial to realize this vision. 
This experiment: i) Utilizes diverse Digital Coherent Optical (DCO) channels with varying data rates, modulation formats and bandwidth, sourced from different system vendors, together with Analog Radio-over-Fiber~(ARoF) transmission to deliver radio services at mmWave (60 GHz) and sub-THz (210 GHz) frequencies, using optical heterodyning, with a shared carrier supplied by the Central Office (CO); ii) Multiplexes the ARoF modulated and unmodulated carriers for heterodyning and DCOs in the same spectral window to maximize the spectral efficiency of the ROADM channels and transparently delivers radio services from a Metro Network to a Passive Optical Network (PON); iii) Demonstrates heterogeneous, converged, multi-vendor coexistence over the live production HEAnet metro network, the Irish research and education network, using a 400 GHz optical bandwidth via commercial OSaaS.


 A field trial of 60 GHz ARoF signals was recently demonstrated \cite{steeg2018public}, providing multiple services like Wi-Fi, 5G and LTE in a mall in Warsaw, Poland. Flexible WDM-based delivery of DRoF, ARoF and SDoF was also recently shown in a lab environment \cite{toumasis2023first}. However, these experiments did not provide metro-access convergence and did not combine DCO and ARoF signals. 
The coexistence of DCO and ARoF over ROADMs was demonstrated in \cite{Ruggeri:22}, using dedicated fibre to distribute a mmWave ARoF signal over the access network. However, this approach implied the use of dedicated access fibers and were laboratory experiments. Recent work \cite{zehao_jlt} demonstrated the coexistence of ARoF, 400 GbE coherent and distributed acoustic sensing over a 32 km field trial in the COSMOS testbed, however, the ARoF signal was at microwave frequency with 100 MHz bandwidth. 

Our experiment successfully demonstrates the coexistence of two ARoF and four DCO channels at rates of 200G, 400G, and 600G over a live production network with 6 multi-degree ROADMs, spanning a 77 km distance. The ARoF mmWave and sub-THz signals, after traversing the metro network, deliver data rates of 5.8 Gb/s and 4 Gb/s, respectively, over a PON topology with a 25 km length and equivalent loss of a 16-way split, achieving pre-FEC BERs below HD-FEC.
\vspace{-2mm}

\section{Experimental Setup}
\vspace{-2mm}

\begin{figure*}[t!]
    \centering
    \includegraphics[width=1\textwidth]{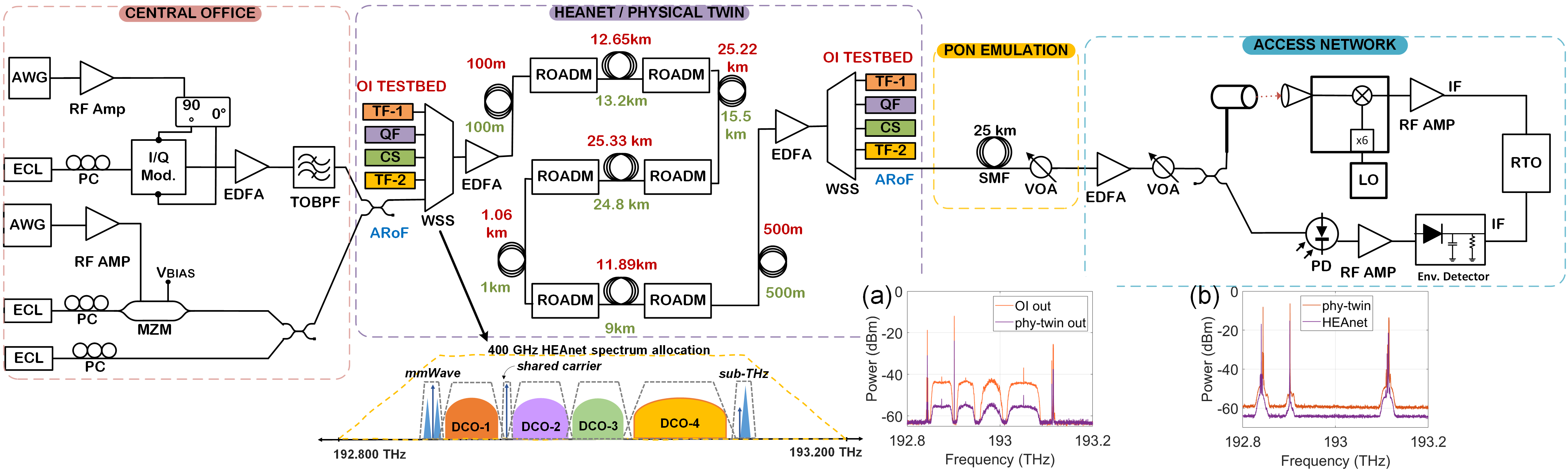}
    \vspace{-8mm}
    \caption{\small Experiment setup, across HEAnet and physical twin, showing ARoF mmWave and sub-THz links. SMF spool lengths in green (red) are for HEAnet (physical twin). (a) Transmitted and received optical spectra going into physical twin consisting of ARoF mmWave and sub-THz carriers and DCO signals, (b) comparison between the ARoF demux outputs of physical twin and HEAnet}
    \label{fig:Arof testbed}
    \vspace{-10mm}
\end{figure*}

\setlength{\tabcolsep}{1.8pt}
\renewcommand{\arraystretch}{1}
\newcolumntype{s}{>{\columncolor[HTML]{AAACED}} p{cm}}
\begin{table*}[t]
\footnotesize
    \centering

    \caption{\small Transmitted Heterogeneuos Signal Parameters} \label{tab:sig_parameters}
    \vspace{-3mm}
    \begin{tabular}{|c|c|c|c|c|c|c|c|c|}
    \hline  
       Parameters            &  mmWave-1& mmWave-2  & sub-THz-1 & sub-THz-2     & DCO-1     & DCO-2     & DCO-3     & DCO-4                \\
        \hline  
        Frequency (THz)      & 192.843  & 192.843   &193.111 & 193.111  & 192.875   & 192.925   & 192.980   & 193.050               \\
        Bitrate ( Gb/s)       & 2     &  6       & 2   & 4        & 400       & 200       & 200       & 600              \\
        Modulation           &16QAM     & 16QAM     & QPSK  & 16QAM     & DP-64QAM  & DP-16QAM  & DP-16QAM  & DP-64QAM             \\
        OFDM Bandwidth (GHz) & 0.6     & 1.5      & 1     & 1         & n/a       & n/a       & n/a       & n/a               \\
        WSS Bandwidth (GHz)  & 18.75       &   18.75      &   25  &   25      & 68.75     & 50        & 50        &   87.5            \\
        IF (GHz)             & 1        & 2         & 3     & 3         & n/a         & n/a         & n/a         & n/a                \\
        \hline  
    \end{tabular}
    \vspace{-6mm}
\end{table*}%

This work combines three systems: a live production metro network, the OpenIreland (OI) \cite{OpenIreland} testbed and the ARoF testbed. The experimental setup is illustrated in Fig. \ref{fig:Arof testbed}. 
The HEAnet production network, shown in Fig.\ref{fig:heanet}, comprises 6 Adtran multi-degree ROADMs, connected by fibre spans totaling 77 km. The experiment was carried out through an OSaaS window of 400 GHz bandwidth (from 192.8 to 193.2 THz), starting at ROADM TCD-1, traversing 4 ROADMs (PW-2, CWT-2, UCD-1 and UCD-2 in Fig.\ref{fig:heanet}) in the metro network, and then exiting at ROADM TCD-2. The 400 GHz window was provisioned through a GMPLS control plane, using the Adtran Optical Ensemble Controller, which also automatically equalises the power levels at every ROADM. Within the system, the OSaaS 400 GHz window is treated as one channel for equalization purposes, thus requiring an offset of 6dB higher power in the ECN.

The OpenIreland testbed has a dual purpose. In the production network experiment, it acts as add and drop nodes for the DCO and ARoF signals. 
One part of OpenIreland acts as the CO node (see Fig.\ref{fig:heanet}), with edge cloud capability, handling the mmWave and sub-THz radio systems and also generating the DCO signals. These have different rates, bandwidth and modulation formats (shown in Table \ref{tab:sig_parameters}), originating from three systems from two vendors. The DCO-1 (400 Gb/s) and DCO-4 (600 Gb/s) are Adtran Teraflex (TF) modules, DCO-2 (200 Gb/s) is an Adtran Quadflex (QF) module, and DCO-3 (200 Gb/s) is an Edgecore Cassini (CS) as labeled in Fig. \ref{fig:Arof testbed}. 
The other part of OpenIreland acts as the edge node where the DCO signals are dropped, and the ARoF signals are sent over the PON and then terminated.
The second purpose of the OpenIreland testbed is to create a physical twin of the HEAnet production network, which is used to estimate channel performance before running the experiment over the HEAnet production network. The physical twin was created by using the same number of ROADMs, but from different manufacturers (Lumentum), loosely matching fiber distance (depending on fiber spool availability in the lab). 
The DCO and ARoF setup remains consistent for both the HEAnet production network and the lab-based physical twin experiments. 

The generation of mmWave and sub-THz ARoF signals at the CO, and their detection at the RU in the access network, are depicted in Fig. \ref{fig:Arof testbed}. The two ARoF signals with different bandwidths and modulation formats (see Table \ref{tab:sig_parameters}) and the shared unmodulated carrier are combined using a coupler and added to the ROADM in OpenIreland, as shown in Fig. \ref{fig:Arof testbed}. The DCO signals are added at the same ROADM. 
After traversing the production metro network, the signals are demultiplexed by the drop ROADM in OpenIreland. 
All DCO signals are dropped back to the respective receiver ports, and the ARoF signals are transparently injected into the PON, consisting of a 25 km SMF and 14.5 dB attenuation representing a 1:16 splitter. At the end of the PON, at the receiver side, the ARoF signals are amplified by an EDFA and then sent to the mmWave or sub-THz receivers. The optical spectra of the transmitted and received signals is depicted in Fig. \ref{fig:Arof testbed}, along with their corresponding ROADM windows (Fig. \ref{fig:Arof testbed} middle bottom).


\textit{mmWave subsystem:}
The 60 GHz mmWave signal is generated using optical heterodyning of an ARoF signal and an unmodulated optical carrier. A double sideband ARoF data signal is generated at the transmitter by modulating an intermediate frequency (IF) OFDM waveform, on an optical carrier using a Mach Zehnder modulator (MZM). An arbitrary waveform generator (AWG) was used to generate the OFDM signal with parameters shown in Table \ref{tab:sig_parameters}. This modulated signal is then coupled with other signals for transmission over HEAnet. At the receiver, the ARoF signal and unmodulated carrier beat on a high-speed photodetector (PD) to generate a double sideband mmWave signal at 60 GHz. An envelope detector detected this signal and downconverted it to IF. A Real-Time Oscilloscope (RTO) captures the IF signal and error performance is evaluated through post-processing.


\textit{Sub-THz subsystem:}
The sub-THz signal is also generated using the optical heterodyne technique. At the transmitter side, the AWG-generated 1 GHz bandwidth OFDM IF signal (parameters shown in Table \ref{tab:sig_parameters}) is modulated on an optical carrier. A 90$^{\circ}$ hybrid coupler and a null-biased IQ-MZM are used to generate a single-sideband optical data signal. 
At the receiver side, the sub-THz ARoF data and un-modulated optical carrier on a waveguide beats at the integrated photodiode antenna 
and generates the OFDM sub-THz signal at 210 GHz. The WIN-PDA THz transmitter, consisting of an InP-based PD and integrated bow-tie antenna, emits the generated signal. A sub-THz receiver downconverts this signal to IF frequency using a waveguide mixer and a local oscillator (LO) generated by a 6x frequency multiplier, as illustrated in Fig. \ref{fig:Arof testbed} receiver setup. The IF signal is captured using an RTO and processed offline to correct for laser frequency drift and phase noise. 
\vspace{-2mm}
\section{Results and Discussion}
\begin{figure}

\begin{minipage}[c]{0.47\linewidth}
\includegraphics[width=0.9\linewidth]{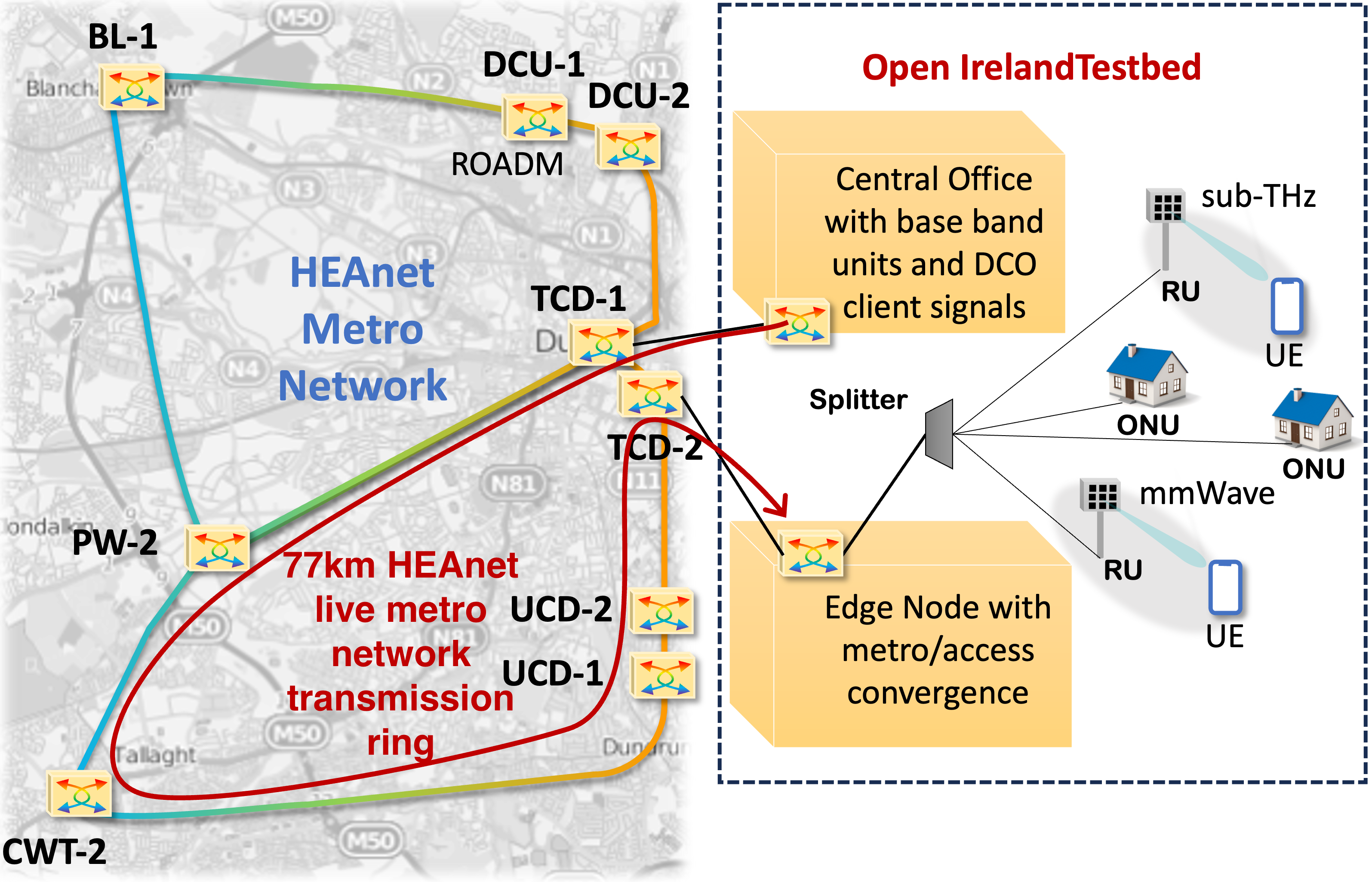} 
    \vspace{-4mm}
    \caption{\small High-level topology of the experiment over the OpenIreland testbed and the HEAnet production network}
    \vspace{-9mm}
    \label{fig:heanet}
\end{minipage}
\hfill
\begin{minipage}[c]{0.52\linewidth}
\includegraphics[width=1\linewidth]{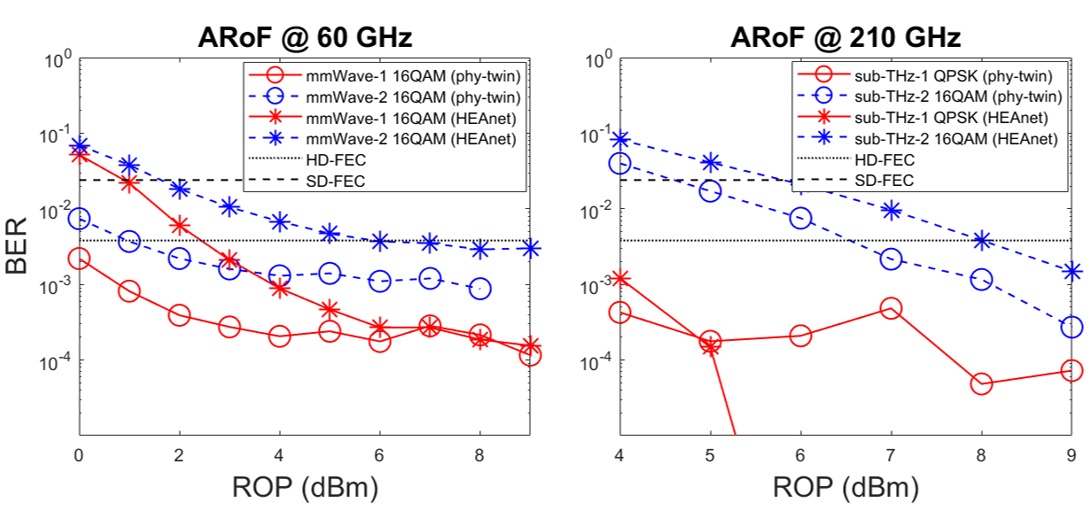}
    \caption{ \small BER curves of mmWave and sub-THz ARoF over converged metro-access with 25 km 16-way PON}
    \vspace{-9mm}
    \label{fig:result 1}
    \end{minipage}
\end{figure}

\vspace{-1mm}
The performance of the DCO and ARoF signals are analyzed through Q-factor and bit error rates (BER), respectively. \uline{Coherent channels are successfully received: DCOs' 1, 2, 3 and 4 signals achieve Q-factor performances of 6, 8.6, 3.7 and 6, respectively, over the physical twin and 6, 8.2, 2.45 and 6, respectively, over the production network.} The low Q-Factor of 2.45 in the Cassini achieves a BER of $7 \times 10^{-3}$, thus still below the SD-FEC threshold. Additionally, despite their spectral adjacency, the ARoF signal did not affect the QoT of the DCO signals. In Fig. \ref{fig:result 1}, both the 16 QAM OFDM mmWave signals, narrow and wide band (refer to Table. \ref{tab:sig_parameters}), attain BER values below the hard-decision forward-error correction (HD-FEC) limit of \(3.8 \times 10^{-3}\), for a received optical power ranging from 1 to 8 dBm for transmission over the physical twin. The wider bandwidth signal suffers slightly more due to the power penalties of the network introduced by the losses of the ROADM's wavelength selective switch (WSS) ($\sim 5 dB$) and the amplifier spontaneous emission noise (ASE) from EDFAs. The fiber launch powers were set to 0 dBm to avoid nonlinearities in the physical twin network. 
Receiver sensitivity at the HD-FEC limit was reduced by 3 dB and 5 dB for narrow- and wideband 16QAM OFDM mmWave signals, respectively, compared to the physical twin network. Despite this, all BER performance remained below the soft-decision FEC limit (SD-FEC) of \(2.4\times 10^{-2}\) up to 1 dBm ROP. The 60 GHz signals delivered data rates of 2.34 and 5.8   Gb/s, for the narrow and wideband signals, respectively.
The sub-THz QPSK and 16QAM OFDM ARoF signals, with data rates of 2 Gb/s and 4 Gb/s, exhibit performance below the HD-FEC limit for powers above 4 dBm and 7 dBm, respectively, for transmission over the physical twin (Fig. \ref{fig:result 1}). The lower SNR requirements for QPSK result in better performance (see Fig.\ref{fig:result 1}). The 16QAM sub-THz signal also shows a 1.5 dB higher power penalty for HEAnet transmission than for the physical twin. QPSK shows bit errors close to zero for both networks and shows the waveform as a robust option in this converged network scenario. Additionally, the sub-THz signal did not affect the mmWave ARoF signal, as the mmWave and sub-THz systems are band-limited to 70 GHz and 170-260 GHz, respectively.
\vspace{-3mm}

\section{Conclusions}
\vspace{-2mm}
 We have demonstrated seamless coexistence of multiple DCO and ARoF signals, placed in close spectral proximity over a production metro network using OSaaS. The worked seamlessly with the automatic channel equalization from the operator management system. The mmWave and sub-THz signals, with 16-QAM OFDM 5G-NR modulation, shared the same optical carrier across the 77 km metro and 25 km PON, and were generated through optical heterodyning at the receiver. \uline{This is an important milestone for flexible heterogeneous service provisioning using OSaaS commercial services.} 
 
\vspace{2mm}
{\footnotesize
\fontdimen2\font=1pt
\noindent\textbf{Acknowledgments.}
This work was supported by EU ECO-eNET (SNS JU) No. 10113933, Science Foundation Ireland (SFI) 18/RI/5721, 13/RC/2077 p2, 12/RC/2276 p2 and 22/FFP-A/10598. 
\vspace{-4mm}


\begin{thebibliography}{99} 
\footnotesize
\vspace{-1mm}
\bibitem{5G_sharing} N. Afraz, et al. Evolution of access network sharing and its role in 5G networks. Applied Sciences, Oct. 2019.
\bibitem{zhu2024optical} Y. Zhu and W. Hu. Optical access networks for fixed and mobile applications. JOCN, Feb. 2024.
\bibitem{steeg2018public} M. Steeg, et al. Public field trial of a multi-rat (60 GHz 5G/LTE/WiFi) mobile network. IEEE W.Comms, Oct. 2018.
\bibitem{toumasis2023first} P. Toumasis, et al. First real-time demonstration of a flexible multi-lambda drof/arof/sdof transport for fiber/mmwave ran. ECOC 2023.
\bibitem{Ruggeri:22} E. Ruggeri, et al. Reconfigurable fiber wireless fronthaul with arof and drof co-existence through a Si3N4 roadm for heterogeneous mmwave 5G c-rans., JLT Aug. 2022. 
\bibitem{zehao_jlt}Z. Wang, et al. Field trial of coexistence and simultaneous switching of real-time fiber sensing and coherent 400 GBe in a dense urban environment. JLT Apr. 2024.
\bibitem{OpenIreland} Openireland testbed, funded by science foundation Ireland. [Online]. Available: www.openireland.eu
\end{thebibliography}
\end{document}